\documentstyle[prb,aps]{revtex}
\begin{document}

\draft

\title{
Zero-temperature magnetism in the periodic Anderson model in
the limit of large dimensions}

\author{
Marcelo J. Rozenberg}
\address{
Laboratoire de Physique Statistique de l'Ecole Normale Sup\'{e}rieure \\
24, rue Lhomond, 75231 Paris Cedex 05, France \\
e-mail: marcelo@physique.ens.fr}

\date{\today}

\maketitle

\begin{abstract}
We study the magnetism in the periodic Anderson model
in the limit of large dimensions by
mapping the lattice problem into an equivalent local impurity self-consistent
model.
Through a recently introduced algorithm based on the exact diagonalization
of an effective cluster hamiltonian, we obtain solutions with and without
magnetic order in the half-filled case.
We find the exact AFM-PM phase boundary which is shown to be of $2^{nd}$
order and obeys $\frac{V^2}{U} \approx const.$
We calculate the local staggered moments
and the density of states to gain insights on
the behavior of the AFM state as it evolves from itinerant to
a local-moment magnetic regime
\end{abstract}

\section{Introduction}

The periodic Anderson model (PAM)
consists of a band of $d-$electrons that
hybridizes with localized $f-$electron states at each lattice site. The double
occupation of the $f-$sites is disfavored by a local
term that corresponds to the repulsive Coulomb interaction.

This model hamiltonian
is widely considered to be relevant for the
description of a large class of strongly correlated systems, most notably,
the so called heavy fermions and Kondo insulators.
Examples of these
systems are, among others, $UPt_3$, $CeAl_3$, the insulators $Ce_3Bi_4Pt_3$,
$CeNiSn$, and possibly
$FeSi$, and $SmB_6$.\cite{lee,fisk}

Despite much effort,
our present theoretical knowledge about
the variety of behaviors that solutions of this model may have, is
still limited.
Much of our current understanding of the model
comes from approaches based in the consideration of a variational
ansatz \cite{riceueda,varma},
and also from its formulation in limiting cases where
the problem becomes more tractable, as for instance,
the large $N$ approach.
\cite{piers,riseborough,millislee}

On the other hand, the steady increase of
computational power, has recently made possible the numerical
study of the model in 2-dimensional finite lattices by quantum Monte Carlo
techniques.
However, that numerical technique
has some intrinsic limitations in regard
of the parameter space that can be investigated, in particular the
$T \rightarrow 0$ limit. \cite{vekic}

In this paper, we will consider the problem within
the local impurity self-consistent approximation (LISA).
This method was introduced in the present
form by Georges and Kotliar \cite{geko}
in the context of the Hubbard model, and later generalized for other
models (including the PAM) by Georges, Kotliar and Si. \cite{gks}
It can be thought of as a dynamical mean field theory for correlated electron
systems.
Its basic idea is to isolate
a local site of the lattice model and embed it in an effective bath
that contains the information of the
response of the rest of the lattice.

The LISA method becomes exact in the limit of large number
of spatial dimensions.\cite{mv} Therefore,
it also represents
a limiting formulation of the model
where simplifications occur (e.g. the self-energy becomes local),
that allow us to obtain essentially exact results.
This method has already been used in the periodic Anderson
model for the investigation of the paramagnetic solution at finite
temperature. \cite{jarrel}

It should be emphasized here, that due to the local character of the
approach, it should be most suitable for the study of electronic systems
where the relevant orbitals are very localized. Also, since the interactions
can be considered in a non perturbative manner, it
is most appropriate when the
correlations play a crucial role. These features are clearly a characteristic
of experimental systems such as heavy fermions and Kondo insulators
where the PAM is considered to be most relevant.

In this paper we focus on the study of solutions with
magnetic order at $T=0$. In particular, we will present the exact
phase diagram. We will also study the nature
of the different antiferromagnetic regimes that the solution displays.
To this end we obtain the staggered magnetic moments and the densities
of states in various regions of the model parameter space.

The question of the study of magnetically ordered solutions was previously
considered by Doniach \cite{doniach} in the
context of the Kondo lattice, and by Evans \cite{evans} and , more
recently, by Sun {\it et al.}
\cite{sun} in the PAM.

We conclude with a brief discussion on the relevance
that our results may have
for the interpretation of experimental results.

\section{Methodology}
The formulation of the LISA method
has been already discussed in detail elsewhere, \cite{gks}
so here we shall
limit ourselves to a brief presentation of the relevant expressions in
particular we shall note how they are modified to consider solutions with
magnetic order.

We begin by writing the action of the lattice
model in its Function Integral form

\begin{equation}
{\cal S}=
- \sum_{k,\sigma} \int^{\beta}_{0} \int^{\beta}_{0} d\tau d\tau' \
\psi^{\dagger}_{k\sigma}(\tau) G_{0\sigma}^{-1}(\tau-\tau')
\psi_{k\sigma}(\tau')
\ + \ U  \sum_{i} \int^{\beta}_{0}d\tau \ (n_{if\uparrow}(\tau)-{1\over2})
(n_{if\downarrow}(\tau)-{1\over2})
\label{action}
\end{equation}

where the annihilation operator is defined as
$\psi_{k\sigma} \equiv \{f_\sigma,d_{k\sigma}\}$, $U$ is the
local repulsion
for double occupation of the $i^{th}$ $f-$site, and $k$ runs over the
Brillouin zone.
The inverse matrix propagator $G_{0}^{-1}$
is explicitly

\begin{equation}
G_{0\sigma}^{-1}(k,i\omega)= \left(
\begin{array}{cc}
i\omega+\mu+\epsilon_f^0 & V_k \\
V_k & i\omega+\mu -\epsilon_k
\end{array}
\right)
\end{equation}

where $\epsilon_f^0$ is the atomic energy of the $f-$site, $\mu$ is the
chemical potential, and $V_k$ is the hybridization matrix element between
the $f$ and the $d-$sites.
Here, for simplicity, we take $V_k=V$, and consider the particle-hole
symmetric case setting
$\epsilon_f^0=\mu=0$.
Since in the limit of $d \rightarrow \infty$ the self-energy becomes
local, i.e. $k-$independent, we can write the inverse Green function as

\begin{equation}
G_{\sigma}^{-1}(k,i\omega)= \left(
\begin{array}{cc}
i\omega -\Sigma_{\sigma}(i\omega) & V \\
V & i\omega -\epsilon_k
\end{array}
\right)
\end{equation}

To proceed further we consider the model on two sublattices A and B as is
usually the case when dealing with antiferromagnetically ordered states,
and demand
the self-consistency condition that follows from the fact that
the local Green function obeys

\begin{equation}
G(i\omega)=
\frac{1}{N}\sum_k G(k,i\omega)=\int_{-\infty}^{\infty}\rho^0(\epsilon)
G(\epsilon,i\omega)d\epsilon
\end{equation}

We consider, for simplicity, the case of a Bethe
lattice, which in the large dimensional limit has a semi-circular
free density of states $\rho^0(\epsilon)={2\over{\pi D}}\sqrt{1-
({\epsilon \over D})^2}$. The half-bandwidth $D$ corresponds to a hopping
parameter $t$ between neighboring $d-$sites with $D=2t$\cite{scale}.
We set $D$ equal
to unity. This density of states has the further advantage that naturally
incorporates the same band-edge behavior as a $3-$dimensional model. For
a discussion of the hypercubic lattice see Ref.\ \onlinecite{jarrel}.

As a result, we arrive at the following {\it local}
effective action

\begin{equation}
{\cal S}_{local}=
- \sum_{\sigma} \int^{\beta}_{0} \int^{\beta}_{0} d\tau d\tau' \
\psi^{\dagger}_{\sigma}(\tau) {\cal G}_{0}^{-1}(\tau-\tau')
\psi_{\sigma}(\tau')
\ + \  U  \int^{\beta}_{0}d\tau \ (n_{f\uparrow}(\tau)-{1\over2})
(n_{f\downarrow}(\tau)-{1\over2})
\label{localaction}
\end{equation}

where $\psi_\sigma \equiv \{f_\sigma,d_\sigma\}$ is the operator associated
to a site that we call origin (of a given sublattice, say, $A$).
The local inverse propagator ${\cal G}_0^{-1}$ has the
explicit form

\begin{equation}
{\cal G}_{0\sigma A}^{-1}(i\omega)= \left(
\begin{array}{cc}
i\omega  & V \\
V & \ \ i\omega-{D^2\over4}[\tilde G_{\sigma B}]_{dd}(i\omega)
\end{array}
\right)
\label{s1}
\end{equation}

where $\tilde G$ is the ``cavity'' Green function
which has the information of the response of the lattice.
In the present case of a Bethe lattice
it simply becomes $\tilde G = G$.
The symmetry properties
of the two sublattice representation imply

\begin{equation}
G_{\sigma A} = G_{-\sigma B}
\label{sym}
\end{equation}

so the self-consistency condition can be more concisely rewritten as

\begin{equation}
[{\cal G}_{0\sigma A}^{-1}]_{dd}(i\omega) =
i\omega-{D^2\over4}[G_{-\sigma A}]_{dd}(i\omega)
\label{s2}
\end{equation}

and the lattice index can be dropped.

At this point it is worth pointing out that ${\cal G}_{0}$ should not be
confused with $G_{0}$. While the latter is the free propagator
of the {\it lattice} model, the former corresponds to the free propagator
of the effective impurity model which is {\it local}. Furthermore,
${\cal G}_{0}$ is a quantity that is unknown {\it a priori}, and that has to be
solved for self-consistently from equations
(\ref{localaction},\ref{s1},\ref{s2}).

We solve this system of equations numerically  by a recently introduced
exact diagonalization procedure.\cite{srkr,rmk,ck}
It basically consist in introducing
a parametrization for the $G_{dd}$ in terms of the parameters of two
continued fraction expansions that describe the ``particle'' ($\omega>0$) and
``hole'' ($\omega<0$) excitations.

\begin{equation}
G_{dd}(\omega)= G_{dd}^{>}(\omega) + G_{dd}^{<}(\omega),
\end{equation}

with

\begin{eqnarray}
G_{dd}^>(\omega)=
\frac
{\langle gs|d d^\dagger| gs\rangle }
{\omega -a_0^> -\frac{b_1^{>2}}{\omega  -a_1^{>}
-\frac{b_2^{>2}}{\omega  -a_2^>-...}}} \nonumber
\\
\nonumber
\\
G_{dd}^<(\omega)=
\frac
{\langle gs|d^\dagger d| gs\rangle }
{\omega -a_0^< -\frac{b_1^{<2}}{\omega  -a_1^{<}
-\frac{b_2^{<2}}{\omega  -a_2^<-...}}}
\end{eqnarray}

The continued fractions can be then thought
of as resulting from an effective electronic bath where the
impurity site is embedded. This bath consists of two chains, one for
each continued fraction. The hopping elements between sites
of the chains are the parameters $b_i^{>/<}$, and the atomic
energies of the sites are the $a_i^{>/<}$.
In Fig.\ \ref{fig1} we schematically show the
resulting effective hamiltonian model for
the local impurity site plus the effective electron bath.
It can be straightforwardly verified that it renders the
self-consistency equation (\ref{s2}).

Swithching on the interaction $U$, the cluster of $N_S$ sites
is then exactly diagonalized and the local
Green functions are calculated as continued fractions.
The parameters of this newly obtained continued fractions, are then
feed back as new parameters for the effective bath chains.
It should be noted that in the present case, where solutions with
antiferromagnetic order are allowed, one should connect the
spin $\sigma$ chains to the spin $-\sigma$ local $d-$site as follows from
equation (\ref{s2}).

The self-consistency is thus translated
into the self-consistent determination of the parameters of a
continued fraction representation
of $G_{dd}$.
This numerical procedure represents an essentially exact solution of the
model. The only approximation consists in the truncation of the length
of the continued fractions, due to the finite size of the effective electron
bath that can be dealt with. It has been demonstrated elsewhere
that this approach is in excellent agreement
with the solution obtained from a quantum Monte Carlo calculation.\cite{srkr}
Moreover, it
allows the investigation of regions of
parameter space that are inaccessible for QMC,
for instance, the $T\rightarrow0$ and large $U$ limit.

\section{Magnetic Order}
\label{magord}
\subsection{Phase Diagram}

We have performed extensive calculations for effective hamiltonian
clusters of $N_S=$ 4, 6, and 8
sites in the ($U,V$) parameter plane for fix $D=1$.
In Fig.\ \ref{fig2} we show the
exact magnetic phase diagram of the model.
The parameter space is divided
into two regions, one paramagnetic and the other with long range
antiferromagnetic order. As is shown in the inset, the boundary line
has the approximate form $U_c \propto V_c^2$. The
interpretation of this result is simple. The onset of magnetic order is
controlled by the crossing of two energy scales that depend on a
single parameter $J$ proportional to the exchange constant
${V^2 \over U}$. If ${J \over D}>>1$, the magnetic moment on the $f-$site
is Kondo-quenched by the band of $d-$electrons, rendering a paramagnetic
state. On the other hand, for ${J \over D}<<1$, the  RKKY interaction
($\sim({D \over J})^\alpha,\ \alpha > 1$)
becomes dominant over the Kondo
effect ($\sim e^{-({D \over J})}$). Therefore, the screening of the
local moment becomes incomplete and the residual magnetic interaction along
with the bipartite nature of the lattice
drives the system to an antiferromagnetic state.
We thus see that these energy scales cross at a critical value $J_c$,
and in consequence the ($U,V$) parameter space is split into two phases
with a boundary that obeys ${V^2\over U} \approx J_c$.
We find from the numerical
solution that $J_c \approx 0.075D$. This
result is consistent with Doniach's estimate for the critical value
of $J_c$. In the context of a one dimensional Kondo lattice \cite{doniach},
he found that at the mean field level $\frac{J_c}{D} \sim {\cal O}(1)$
and argued that fluctuation effects should strongly reduce the ratio.

It is interesting to note that this result is qualitatively similar to
the recently obtained by Veki\'{c} {\it et al.}
for a 2-dimensional finite size
lattice using quantum Monte Carlo.\cite{vekic}
This lack of dimensional dependence represents a strong evidence
for the applicability of the LISA method in the study of electronic systems
where localized states and correlations play an essential role.

As it turns out, the transition is of $2^{nd}$ order along the whole
critical line. This will be illustrated by the study
of the magnetization below.

\subsection{Staggered Magnetization}
We now consider the evaluation of the staggered magnetization.
Since the present effective cluster
hamiltonian method treats explicitly the $f$ and $d-$sites,
their local magnetization can be simply and efficiently evaluated as
expectation values on the ground state.

\begin{equation}
m_{Zf} = \langle gs| n_{f\uparrow}-n_{f\downarrow}|gs \rangle, \ \ \
m_{Zd} = \langle gs| n_{d\uparrow}-n_{d\downarrow}|gs \rangle
\end{equation}

In Fig.\ \ref{fig3}, we show the magnetic moments as a function of the
hybridization $V$
for different values of the interaction $U$.
Both moments $m_{Zd}$ and $m_{Zf}$ vanish continuously at the transition
for any value of $U$ indicating that the
transition is of $2^{nd}$ order along the whole critical line.

A notable feature of the results is that the size of
the magnetic moments of the $f$ and $d-$sites are very different, except
close to the critical point.
We also point out that while the magnetization
of the $f-$site becomes always rapidly saturated at $m_{Zf} \approx 1$,
independently of the
strength of the interaction, the same does not occur on the
$d-$sites.
As $U$ is increased, the maximum of $m_{Zd}$ is enhanced.
This suggests that it would be interesting to consider
a limiting case of the model with both $V,U \rightarrow \infty$
(or equivalently $D \rightarrow 0$) and ${V^2}
\over U$ kept constant (i.e., $\frac{J}{D} \rightarrow \infty$).
This model may present an almost compensated
total magnetic moment.

An important remark is that the deep Kondo regime, with an exponentially
small gap in the paramagnetic density of states, \cite{riceueda,piers,jarrel}
is actually hindered by the onset of the magnetic instability.
This is relevant for the understanding of the very small gaps
observed in the Kondo insulators.
This situation could,
in principle, be modified by introducing {\it frustration} in the model, as
for instance through $n.n.n.$ hopping \cite{rkz},
or by the consideration of degeneracy
in the $f-$orbital.

Let us now consider in more detail the behavior of these quantities
in the case where
the critical line  is approached and  also
when $J \rightarrow 0$ (${V^2 \over U} \rightarrow 0$).

In Fig.\ \ref{fig4} we show the magnetic moments as a function of the
hybridization $V$ for a fixed value of the interaction $U=2$ in clusters
of various sizes. A similar result
was obtained by Sun {\it et al.}, using a variational Gutzwiller
approximation.\cite{sun}

As was already pointed out in that paper,
the behavior of the magnetization as $V \rightarrow 0$ and
as $V \rightarrow V_c$, near the phase transition, is quite different.
In the latter case, both $m_{Zf}$ and
$m_{Zd}$ vanish when the Kondo effect becomes relevant
close to the transition line.
In the former case, however, we see that while
the magnetic moment of the $f-$site becomes saturated and
fully developed $m_{Zf} = 1$,
the local moment of the $d-$electron band goes to zero. This result is
demonstrated by
the data shown in the inset, where we plot the magnetization at
at small $V=0.01$ as a function of the size of the cluster.

It is also important to note from the comparison of the results from
clusters of size $N_S =4,\ 6$ and $8$ shown in the main part of the figure,
that although the
value of the magnetization is notably affected by the system size, the
the critical point for the ordering transition remains
basically size independent. This
feature makes the phase diagram presented in the previous section an
essentially exact result for this model.
Also, a scaling $\propto \sqrt{1-\frac{V}{V_C}}$for the moments
$m_{Zd}$ and $m_{Zf}$ is found near the critical line. \cite{sun}
Moreover, we note that the results for $m_{Zf}$, which are
essentially independent of the system size,
can be very well fit in the whole
interval $0 < V < V_c$ with the function $\sqrt{1-{(\frac{V}{V_C})}^3}$
with only $V_c$ as fitting parameter.

We now investigate the behavior of the moment along the other direction in the
($U,V$) plane. In Fig.\ \ref{fig5} we show the results for the magnetization
as a function of $U$ with fixed $V=0.5$. A notable feature is that
a relatively large magnetization on the $d-$sites seems to persist
for extremely high values of
the interaction $U\sim100$ (which are easily accessible within
the present numerical
technique), despite the fact that the RKKY coupling should be very small.
However, as the inset demonstrates,
when the data are plotted against the relevant
energy scale of the model, i.e. $J \sim {{V^2}\over U}$,
the behavior turns out to be
qualitatively similar as in the previous case (which is also an indication
that finite size effects are important in this parameter
region).

To fully elucidate the behavior of the magnetization in the limit
${V^2 \over U} << D < U$ ($J/D \rightarrow 0$), we
obtain the Hartree-Fock solution of the model ({\it c.f.} Appendix).
The approximation
becomes essentially an exact result in this limit, where
correlation effects are less important.
We find that the staggered  magnetization of the $d-$sites behaves
as

\begin{equation}
m_{Zd} \approx \frac{8}{\pi} \frac{J}{D} ln\left( \frac{2D}{J} \right),
\ \ \ \ J << D
\label{mzd}
\end{equation}

where $J\equiv {V^2 \over U}$. This result is plotted for comparison with the
corresponding data from exact diagonalization in Fig.\ \ref{fig4} (dotted
line).

To end this section we would like to briefly consider the behavior
of the double occupation $\langle D \rangle$.
This quantity is simply related to the
square of the magnetization by $\langle m^2\rangle =1-2\langle D\rangle $.
The results for fixed $V$ and as a function of $U$
from exact diagonalization are presented in Fig.\ \ref{fig6}.
For comparison, we also plot similar results for the two-site local
model which corresponds to the limit where the band-width
$D \rightarrow 0$ and
can be solved analytically.
When $U=0$, we have $\langle D\rangle=1/4$
for both the $d$ and $f-$sites since, in the non-interacting
case, the
four spin configurations at each site appear with equal probability.
As $U$ is increased, the singlet state becomes energetically
favorable and the double occupation decreases.
This behavior is  qualitively well captured by the
two-site model, however, as it does not contain the
lattice feedback, it underestimates the value of $\langle D_f\rangle$ and
overestimates $\langle D_d\rangle$.
This difference can be understood from the following arguments:
For the $f-$electrons the lattice contains the RKKY interaction
(that the two-site model lacks) which
favors the development of the local magnetic moment to minimize the
energy. On the other hand, for the $d-$electrons, the enhancement of
the double occupation respect to the two-site model is simply due to their
band-like character which is brought in through the self-consistency condition.

When the interaction $U$ is further increased
the transition into the antiferromagnetic state takes place, as shown
by the results for the staggered magnetization $m_{Zd}$ that
we reproduce in the figure for comparison.
The most notable feature is that $\langle D_d\rangle$
has non-monotonic behavior. It
experiences an upturn and returns to the non-interacting limit value
when $U\rightarrow \infty$.  This behavior merely reflects the
fact that as $U$ increases, the effective magnetic exchange interaction $J$
between the sites becomes small. More precisely, one may
perform a Schrieffer-Wolff transformation \cite{sw}
and realize that $d$ and $f-$sites become gradually decoupled with
the $d-$electrons approaching the non-interacting band limit, as
$J \sim {V^2 \over U} \rightarrow 0$, i.e., when $U \rightarrow \infty$.
On the other hand, for the case of the $f-$electrons, where
the local repulsion is the dominant interaction,
the simple two-site model can still capture the correct qualitative behavior.

\section{Itinerant and localized magnetic behavior}
\label{itloc}
In this section we consider the behavior of the $f-$electrons as
they evolve from the itinerant to the localized magnetic regime.
Different aspects of this question were
first discussed by Doniach \cite{doniach} and Evans \cite{evans},
and later by Sun {\it et al.}\cite{sun}.

The existence of two qualitatively different regimes was already
suggested by
the two routes in which the system looses its magnetic order:
$(i)$ when $V \rightarrow 0$ or $U \rightarrow \infty$ ($J \rightarrow 0$),
the $f-$moments became free local spins with $m_{Zf} \rightarrow 1$
and the $d-$electrons, having $m_{Zd} \rightarrow 0$, approach
a non-interacting band; and $(ii)$ at the $2^{nd}$ order critical line,
as both $m_{Zd}$ and $m_{Zf} \rightarrow 0$.

In what follows, we shall gain new insights on these
two distinct regimes by studying the
redistribution of spectral weight of the electronic density of states.

At $T=0$, the Green functions of the problem can be obtained as
two continued fraction $G_{c\sigma}(\omega) = G_{c\sigma}^{>}(\omega) +
G_{c\sigma}^{<}(\omega)$, with $G_{c\sigma}^{>}$ and $G_{c\sigma}^{<}$
corresponding to ``particle'' and ``hole'' excitations
respectively \cite{srkr,rmk,ck}

\begin{eqnarray}
G_{c\sigma}^{>}(z)= \langle gs| c_\sigma
\frac{1}{z-(H_{eff}-E_{gs})}c^{\dagger}_\sigma|gs \rangle  \nonumber \\
G_{c\sigma}^{<}(z)= \langle gs| c^{\dagger}_\sigma
\frac{1}{z+(H_{eff}-E_{gs})}c_\sigma|gs \rangle
\end{eqnarray}

where the anihilation (creation) operator
$c$ $(c^\dagger)$  generically stands for $d$ $(d^\dagger)$ and
$f$ $(f^\dagger)$, and $H_{eff}$
is the effective cluster hamiltonian introduced before.
The $-\sigma$ Green function is by symmetry
$G_{c{-\sigma}}(\omega)=-[G_{c\sigma}(-\omega)]^*$.

In Fig.\ \ref{fig7} we display the density of states for the $f$ and
$d-$electrons. They are obtained from the imaginary part of the
local Green functions of
clusters with 8 sites. Note that, unlike the quantum Monte Carlo
technique that has also been applied to this problem, our results
are directly obtained on the real axis with no need for analytic
continuation. However, the price we pay is a discrete number of
poles that we broaden in the figures for easier visualization.

The figure illustrates the changes that take place in the density of
states as we move through the different regimes on the phase diagram.
The three sets of curves correspond to the antiferromagnetic phase,
as we move away from the $2^{nd}$ order critical line towards the
small $J$ regime ($V^2 << UD$). The two curves on the top of the figure
correspond to $U=3$ and $V=0.5$ that places the system close to the
critical point. Both spectra are slightly asymmetric which correspond
to the simultaneous vanishing of the magnetic moments that we
discussed before. A notable aspect is that a central feature
can be clearly observed in the $f-$electron spectral function.
This corresponds to the Kondo coherent quasiparticle excitations that are
characteristic in the paramagnetic solution of the model. In this case
it is interesting to observe that the quasiparticle excitations
can survive the onset of magnetic order. In Fig.\ \ref{fig8} we display
the behavior of the $f-$electron density of states in greater detail.
The symmetric curve (grey line) is obtained in the paramagnetic region
very close to phase boundary. This solution corresponds to the
well known scenario of a hybridization band insulating state with the
hybridization amplitude being renormalized by the effect of the
correlations. The low energy features are originated in the splitting of
the Kondo resonance due to the lattice effect.
This scenario
is borne out from many different approaches, as for instance
the Gutzwiller variational wavefunction \cite{riceueda,varma},
large $N$ \cite{piers,riseborough,millislee},
and large dimensions \cite{jarrel}.
We now proceed to slowly move into the antiferromagnetic phase.
The comparison of the paramagnetic solution to the curves obtained close to
phase boundary, but in the antiferromagnetic regime,
provides us new insights as we realize that the transfer of weight
actually occurs at {\it all} energy scales. Both, the high energy features
and the low energy excitations associated with the quasiparticles on the left
side of the spectrum, are
loosing their weight at a similar rate. This interesting behavior
actually makes difficult the task of reducing the the problem to
simpler effective model that could be more tractable.
It also does not allow for a simple interpretation of which
aspect, the itinerant quasiparticles or the localized moments, is the
driving force behind the transition.

We now come back to Fig.\ \ref{fig7} as we decrease the coupling
between sites, setting $U=3$ and $V=0.3$ (central curves). We, thus, move away
from the $2^{nd}$ order line well into the antiferromagnetic regime.
The quasiparticles features have gradually disappeared as the weight
is transferred to the high energy part of the spectrum. This corresponds to the
rapid increase in the size of the $f-$magnetic moment as $V^2$ becomes
much greater than $UD$ and the $f-$electrons localize.

As we finally enter the small $J$ regime, setting $V=0.1$ (lower curves),
we clearly observe how the electron become decoupled. The $f-$electron
spectrum corresponds to a
fully saturated and polarized moment with the totality
of the spectral weight at high energies. The lack of any low frequency
feature indicates that these electron have essentially localized.
On the other hand, the $d-$electron density of states seems to have
closely approached that of a non-interacting band. However, we shall
see next, this is not exactly the case.

Once more we can gain further insights in this limit by using the
Hartree-Fock approximation, which is very accurate as $J << D$ ($V^2 << UD$).
We focus on the $d-$electrons which have a more interesting behavior.
In this case, we can obtain an analytic expression for the density of states
({\it c.f.} Appendix),

\begin{equation}
\rho_d(\omega)= \frac{2}{\pi D^2}
\sqrt{{\left[ \left(\omega-\frac{V^2}{\omega-\frac{U}{2}}\right)
	     \left(\omega-\frac{V^2}{\omega+\frac{U}{2}}\right)\right]}^2
- D^2 {\left[ \left(\omega-\frac{V^2}{\omega-\frac{U}{2}}\right)
             \left(\omega-\frac{V^2}{\omega+\frac{U}{2}}\right)\right]}}
\label{dos}
\end{equation}

We plot this density of states, along with the results from
an 8 site cluster, is Fig.\ \ref{fig9}. It is remarkable to observe
the effort that the few poles from the exact diagonalization method
do to best approximate the essentially
exact continuum result. Even details as the  small weight
at $\omega \approx 1.5$ that originates from the hybridization with
the high energy feature of $\rho_f$ ({\it c.f.}
Fig.\ \ref{fig7}) are well captured.
As expected, the $d-$electron density of states is insulating with
a small gap of size of the order of $J$. However, it is surprisingly
far from regularly approaching the non-interacting limit.  A very sharp
peak develops on the low frequency edge of the gap. When the
$J \rightarrow 0$ limit is approached the peak becomes narrower and
gradually looses its spectral weight.  This feature could, in principle,
be observed in infrared photoemission
spectroscopy.
Also, in transport measurements it may be possible to
observe a strong pressure dependence at temperatures of the size of the gap.

\section{Conclusions}
In this work we have considered solutions with magnetic order
in the periodic Anderson model in the limit of large dimensionality using
the LISA method.
Our results validate and expand on that of Sun {\it et al.} who obtained
approximate solutions based on a variational procedure.

The $T=0$ exact magnetic phase diagram of the model is obtained in detail
by a recently introduced numerical algorithm that is based on the exact
diagonalization of an effective cluster hamiltonian. We find a
paramagnetic-antiferromagnetic $2^{nd}$ order phase boundary that splits
the ($U,V$) parameter space in two regions. The critical line obeys
$U_c \propto V_c^2$ as ${U}$ becomes large.
This is a concrete realization of earlier ideas of
Doniach in the context of the Kondo lattice, and is consistent with recent
numerical results on 2-dimensional finite lattices.

The consistent picture that emerges from this work, very similar to
the two dimensional lattice results, represents
further evidence of the applicability of the LISA method to investigate
the physical behavior of systems that are characterized by the
simultaneous presence of localized orbitals and strong correlations.

{}From the determination of the phase diagram we found that the paramagnetic
Kondo regime with an exponentially small gap, which is relevant for the
description of Kondo insulator compounds, is inaccessible in all parameter
space due to the onset of the antiferromagnetic instability.

The calculation of the magnetic moments in the phase with long range order,
revealed that the possibility of a nearly perfect magnetic spin compensation
that would be relevant for the problem of the ``small moments'', is not
a general feature of the solutions of the model.

An almost perfect compensation of the moments is
only obtained near the $2^{nd}$ order boundary line, where the total
site magnetization
starts to become non-zero as the Kondo screening ceases to be complete.
Therefore, a small moment regime
would require a fine tunning of the parameters
of the model, which is not {\it a priori}
a feature that one expects to find in real systems.
The question weather this situation would persist in the
non particle-hole symmetric case and how the introduction of frustration
may modify the present picture is currently under investigation.

Finally, the method allows to obtain reliable information on the
distribution of the spectral weight in the density of states which
provided interesting new insights on the nature of the phase transition
and that of the different antiferromagnetic regimes.

\acknowledgements
Useful discussions with G. Kotliar, E. Miranda, A. Georges and L. Laloux are
gratefully acknowledged.

\appendix
\section{The Hartree-Fock solution}

We present in this appendix the solution of the self-consistent equations
(\ref{localaction},\ref{s1},\ref{s2}) in the limit ${V^2\over U} << D < U$.
It is a useful exercise that would allow the interested
reader to better grasp some details of the method.
Performing the Hartree-Fock decoupling to interaction term of
the effective local action
(\ref{localaction}), and using that $\langle n_{f_\sigma} - n_{f_{-\sigma}}
\rangle = \langle m_{Zf} \rangle$ and $\langle n_{f_\sigma} + n_{f_{-\sigma}}
\rangle = 1$, the action becomes quadratic
and we can immediately obtain the local Green function for the
$f_\sigma-$electrons.

\begin{equation}
G_{Af_{\pm\sigma}}(\omega)=\frac{1}{\omega \mp m_{Zf}\frac{U}{2} -
V^2 G_{Bd_{\pm\sigma}}(\omega)}
\end{equation}

where $A$ and $B$ denote the sublattice. We now combine this
expression with the symmetry relation between the sublattices
Green functions (\ref{sym}), to obtain

\begin{equation}
G_{Af_{\pm\sigma}}(\omega)=\frac{1}{\omega \mp m_{Zf}\frac{U}{2} -
V^2 G_{Ad_{\mp\sigma}}(\omega)}
\label{a1}
\end{equation}

since the expression involves only quantities on the same sublattice,
the index may be dropped. We now need to find an expression for
the $d-$electron Green function. Once more, from the local action
we obtain,

\begin{equation}
G_{Ad_{\pm\sigma}}(\omega)=\frac{1}{
\omega - \frac{V^2}{\omega \pm {U\over 2}} -
t^2 G_{Bd_{\pm\sigma}}(\omega)}
\end{equation}

where from the same symmetry considerations we have,

\begin{equation}
G_{d_{\pm\sigma}}(\omega)=\frac{1}{
\omega - \frac{V^2}{\omega \pm {U\over 2}} -
t^2 G_{d_{\mp\sigma}}(\omega)}
\end{equation}

This equation is actually two
coupled equations for $G_{d_{\uparrow}}$ and $G_{d_{\downarrow}}$.
We then replace $G_{d_{\sigma}}$ into its same
expression, in order to decouple them.
We obtain,

\begin{equation}
G_{d_{\pm\sigma}}(\omega)=\frac{1}{
\omega - \frac{V^2}{\omega \pm {U\over 2}} -
t^2 \frac{1}{
\omega - \frac{V^2}{\omega \mp {U\over 2}} -
t^2 G_{d_{\pm\sigma}}(\omega)}}
\end{equation}

We are left now with two quadratic equations, each with a single unknown,
which we can solve to obtain the explicit result,

\begin{equation}
G_{d_{\sigma}}(\omega)=
\frac{
\left(\omega-m_{Zf}\frac{U}{2}\right)
\pm\sqrt{
\left(\omega-m_{Zf}\frac{U}{2}\right)^2 - 4t^2
\left(\frac{\omega-m_{Zf}\frac{U}{2}}{\omega+m_{Zf}\frac{U}{2}}\right)}}
{2t^2}
\label{a2}
\end{equation}

from which follows the $d-$electron density of states that we
express in terms of the half-bandwidth $D=2t$

\begin{equation}
\rho_{d_{\sigma}}(\omega)= \frac{2}{\pi D^2}
\sqrt{{\left[ \left(\omega-\frac{V^2}{\omega-m_{Zf}\frac{U}{2}}\right)
	     \left(\omega-\frac{V^2}{\omega+m_{Zf}\frac{U}{2}}\right)\right]}^2
- D^2 {\left[ \left(\omega-\frac{V^2}{\omega-m_{Zf}\frac{U}{2}}\right)
             \left(\omega-\frac{V^2}{\omega+m_{Zf}\frac{U}{2}}\right)\right]}}
\label{a4}
\end{equation}

We still need an equation for $m_{Zf}$. It is simply obtained from
its definition,

\begin{equation}
m_{Zf}= -\frac{1}{\pi}\int_{-\infty}^0 Im \left(G_{f\uparrow}(\omega) -
G_{f\downarrow}(\omega)\right)\
d\omega
\label{a3}
\end{equation}

Thus, (\ref{a1},\ref{a2},\ref{a3}) form a system of equations that have to be
self-consistently solved for.

In general, these equations can be solved numerically. However,
from the results of section \ref{magord} we learn that in the
limit ${V^2\over U} << D < U$, which we are currently concerned with,
we can take $m_{Zf}=1$ to an excellent approximation ({\it c.f.}
Fig.\ \ref{fig4}).
Thus, equation (\ref{a4}) immediately leads to the expression for the
density of states (\ref{dos})
in section \ref{itloc}.
We can finally obtain the behavior of the
magnetization $m_{Zd}$ by replacing
into

\begin{equation}
m_{Zd}= -\frac{1}{\pi}\int_{-\infty}^0 Im \left(G_{d\uparrow}(\omega) -
G_{d\downarrow}(\omega)\right)\
d\omega
\end{equation}

an extracting the leading contribution in this limit ({\it c.f.}
Eq.\ (\ref{mzd}) in section \ref{magord}).

\newpage

\begin{figure}
\caption{
Effective cluster hamiltonian. It consists of a central impurity
site with a $d$ and $f-$electron site connected to an effective electron
bath. $U$ is the on site repulsion of the $f-$site, $V$ is the $d-f$
hybridization matrix element, and $t$ is the $d-d$ hopping amplitude.}
\label{fig1}
\end{figure}

\begin{figure}
\caption{
Phase diagram of the PAM. The inset shows in a $log-log$ plot that
the $2^{nd}$ order critical line obeys $U_c \approx V_c^2$ as $U$ becomes
large. $N_S = 6$.}
\label{fig2}
\end{figure}

\begin{figure}
\caption{
The staggered magnetic moments $m_{Zd}$ and $m_{Zf}$
as a function of the
hybridization $V$ for different values of
the interaction $U=1,5,10$. $N_S = 6$.
}
\label{fig3}
\end{figure}

\begin{figure}
\caption{
The staggered magnetic moments $m_{Zd}$ and $m_{Zf}$
a function of the
hybridization $V$ for a fix value of the interaction $U=2$. For comparison we
plot similar results for $N_S = 6$ (bold lines) and $N_S = 4$ (thin lines).
Note that while the magnitude of $m_Z$ depends on the size of the cluster, the
position of the critical point does not. The dotted line in the top part of
the figure indicates the Hartree-Fock results, while
the one in the bottom is the fit $\protect\sqrt{1-{(\frac{V}{V_C})}^3}$.
The inset shows $m_{Zd}$ and $m_{Zf}$
(top and bottom) for a small value of $V=0.01$ as a function of the inverse
of the cluster size $1/N_S$. $m_{Zd}$ becomes small while $m_{Zf}$ remains
saturated at $m_{Zf}=1$.}
\label{fig4}
\end{figure}

\begin{figure}
\caption{
The staggered magnetic moments $m_{Zd}$ and $m_{Zf}$
as a function of the
interaction $U$ for a fixed value of the hybridization $V=0.5$. $N_S = 6$.
The inset contains the same quantities plotted
as a function of $V^2$ (for fixed $U=5$), and
of $1/U$ (for fixed $V=0.5$). As the
relevant energy scale of the problem is $J \sim V^2/U$, the behavior of the
magnetic moments is qualitatively similar in both cases.}
\label{fig5}
\end{figure}

\begin{figure}
\caption{
Double occupation $\langle D \rangle$ as a function of $U$ for $V=0.5$ from
exact diagonalization of a 6 sites cluster (bold lines).
$\langle D_d \rangle$ is the upper curve and $\langle D_f \rangle$ the lower
one. For comparison we include in dotted line
similar results for a 2 sites cluster
($\langle D_d \rangle$ = $\langle D_f \rangle$). The thin line corresponds
to $m_{Zd}$ that indicates the magnetic regimes.
}
\label{fig6}
\end{figure}

\begin{figure}
\caption{
The local density of
states $\rho_{f_\sigma}, \rho_{d_\sigma}$ (left and right) in different
magnetic regimes.
The top spectra correspond to $V=0.5$ and $U=3$
close to the $2^{nd}$ order critical line. The low energy features of the
$\rho_{\sigma f}$ correspond to quasiparticle AFM. The bottom spectra
correspond to $V=0.1$ and $U=3$. The single peak at a frequency $\approx
\frac{U}{2}$ and the absence of quasiparticles indicate local moment magnetism.
The middle spectra correspond to $V=0.3$ and $U=3$ in the crossover region.
$N_S = 8$.}
\label{fig7}
\end{figure}

\begin{figure}
\caption{
The local density of
states $\rho_{f_\sigma}$ for
$U=2$ and $V=0.45$ in the paramagnetic phase (grey line)
and $V=0.40$, $0.35$ in the antiferromagnetic phase (dotted and full lines).
}
\label{fig8}
\end{figure}

\begin{figure}
\caption{
The local density of
states
$\rho_{d_\sigma}$ for $U=3$ and $V=0.2$. The bold line corresponds to
the Hartree-Fock solution and the thin line to the exact diagonalization
of an 8 sites cluster.
}\label{fig9}
\end{figure}


\begin{references}
\bibitem{scale}
As usual in the large$-d$ limit, we rescale the hopping parameter
$t \rightarrow \frac{t}{\sqrt{q}}$ with $q$ the connectivity of the
lattice.
\end{references}
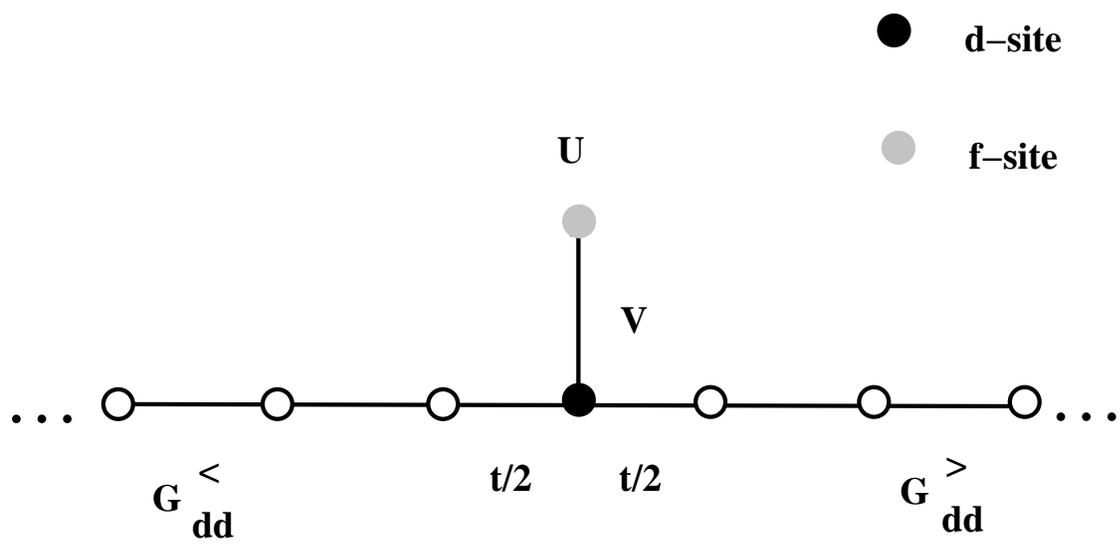
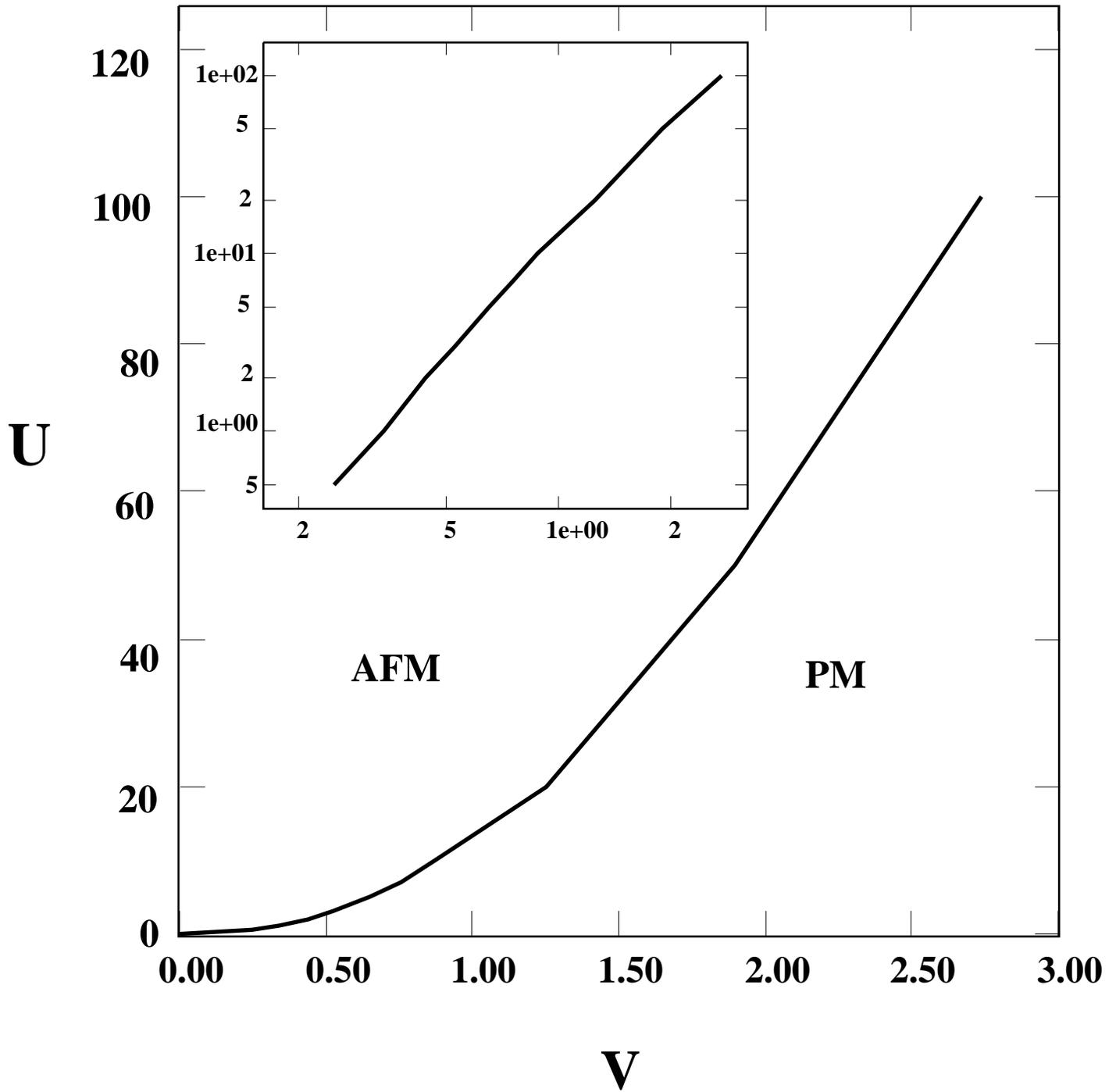
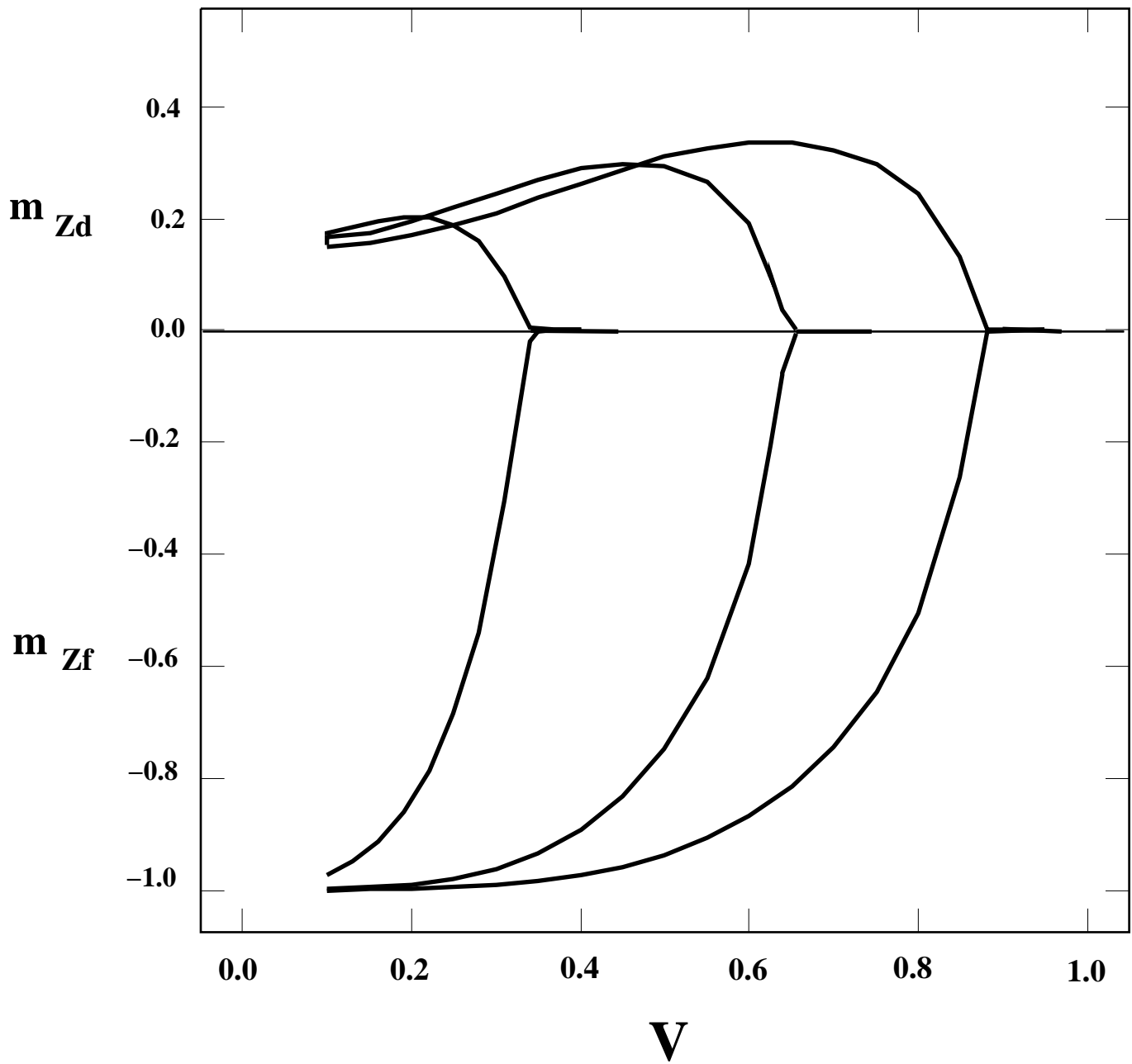
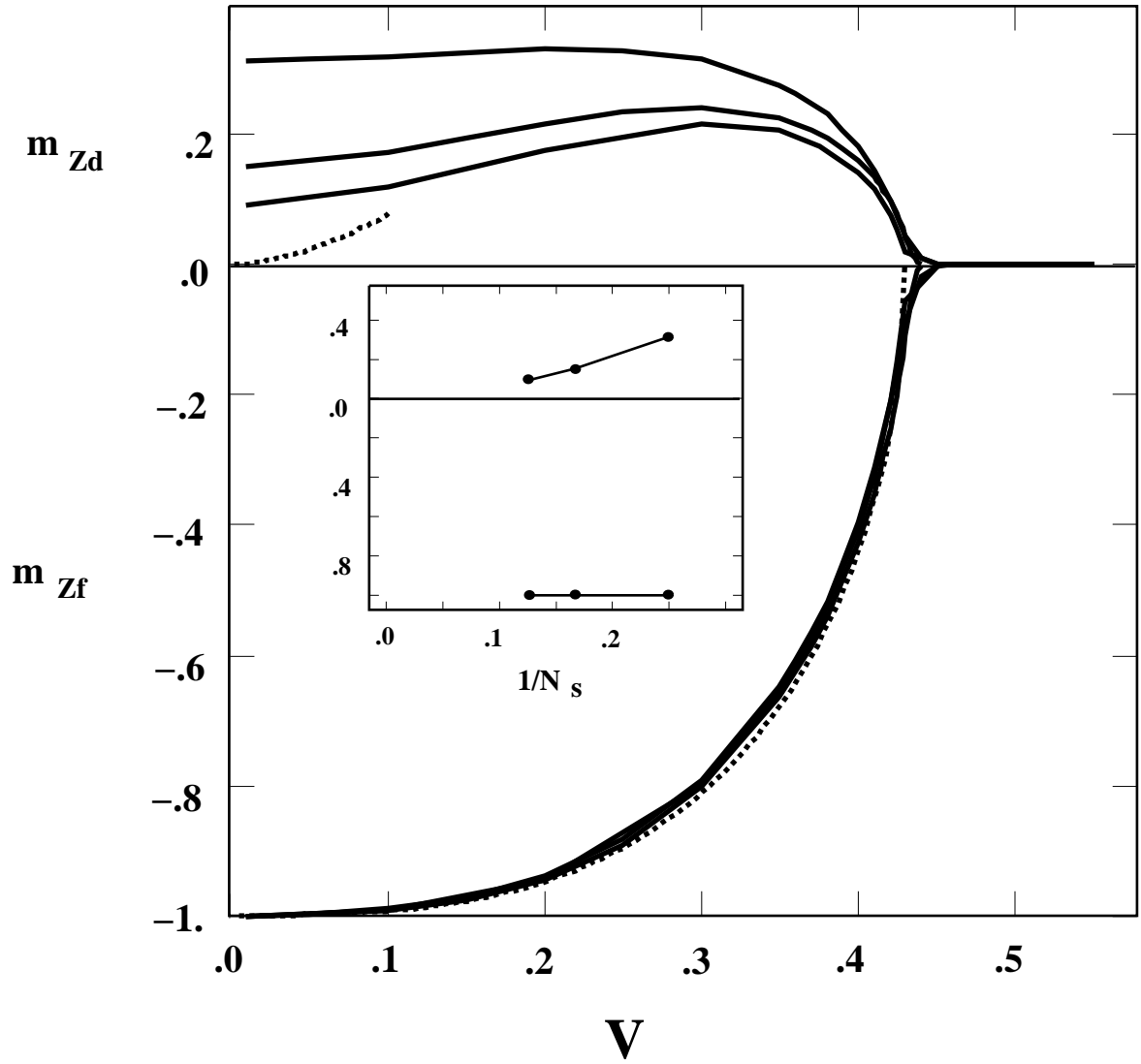
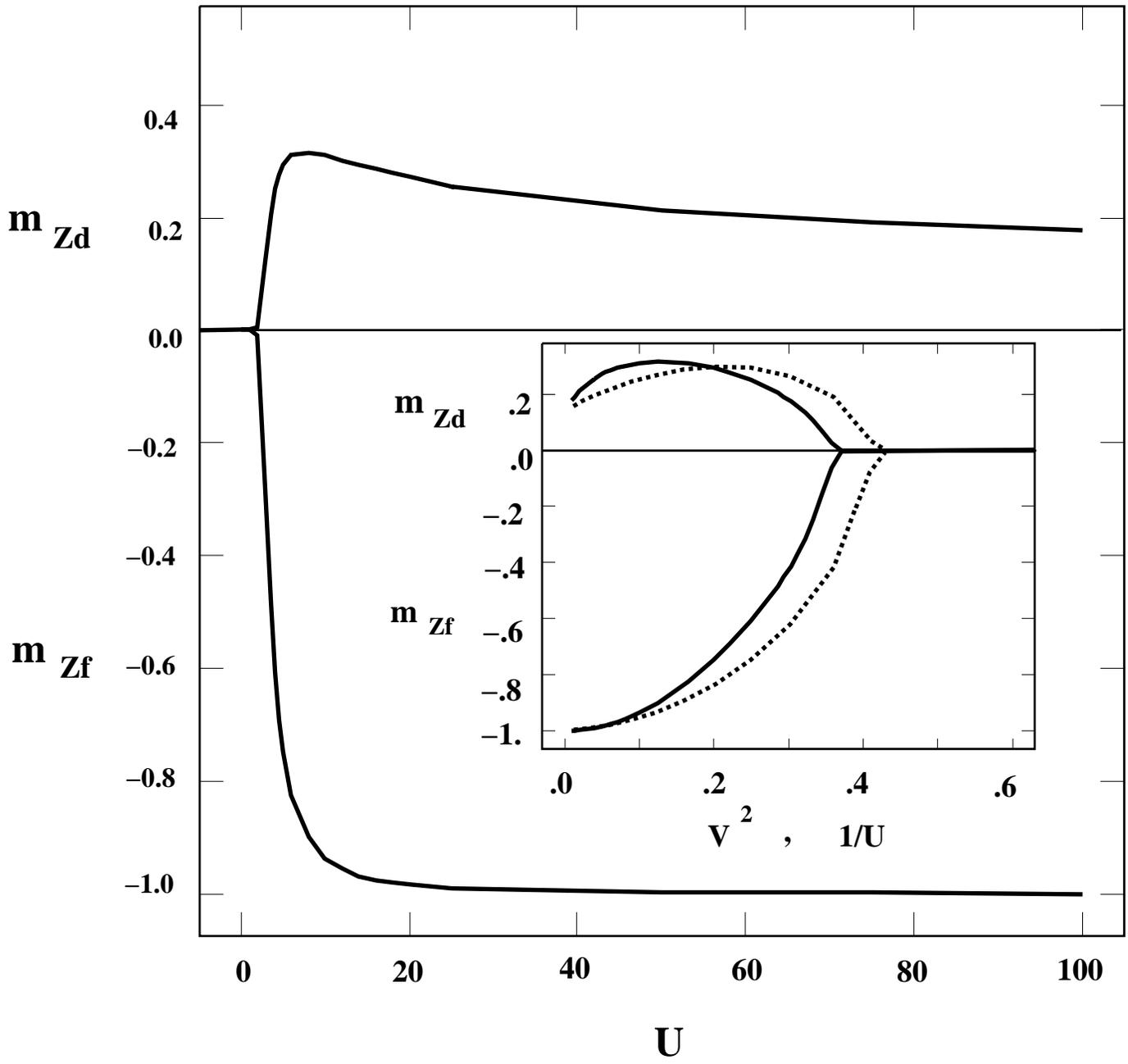
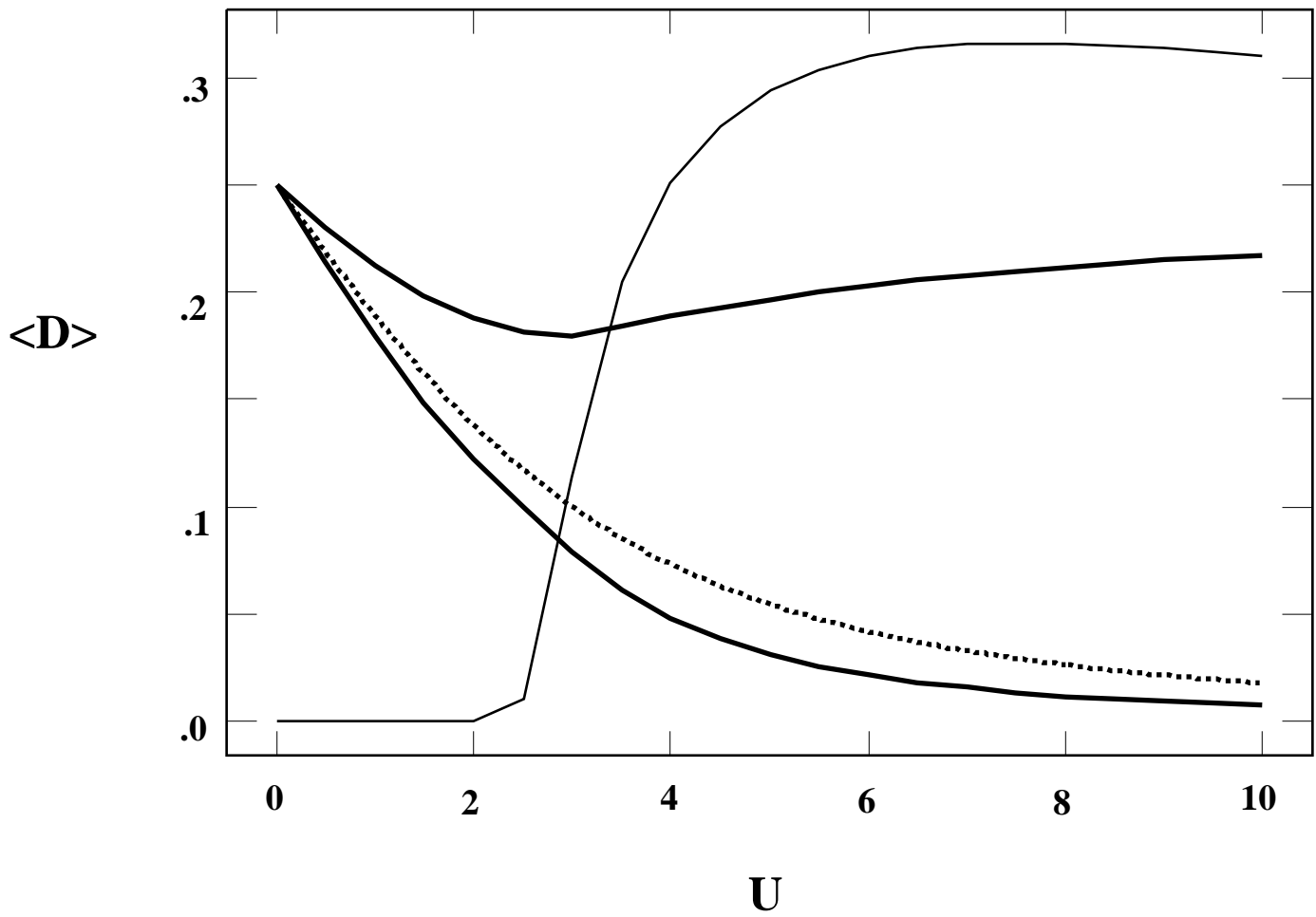
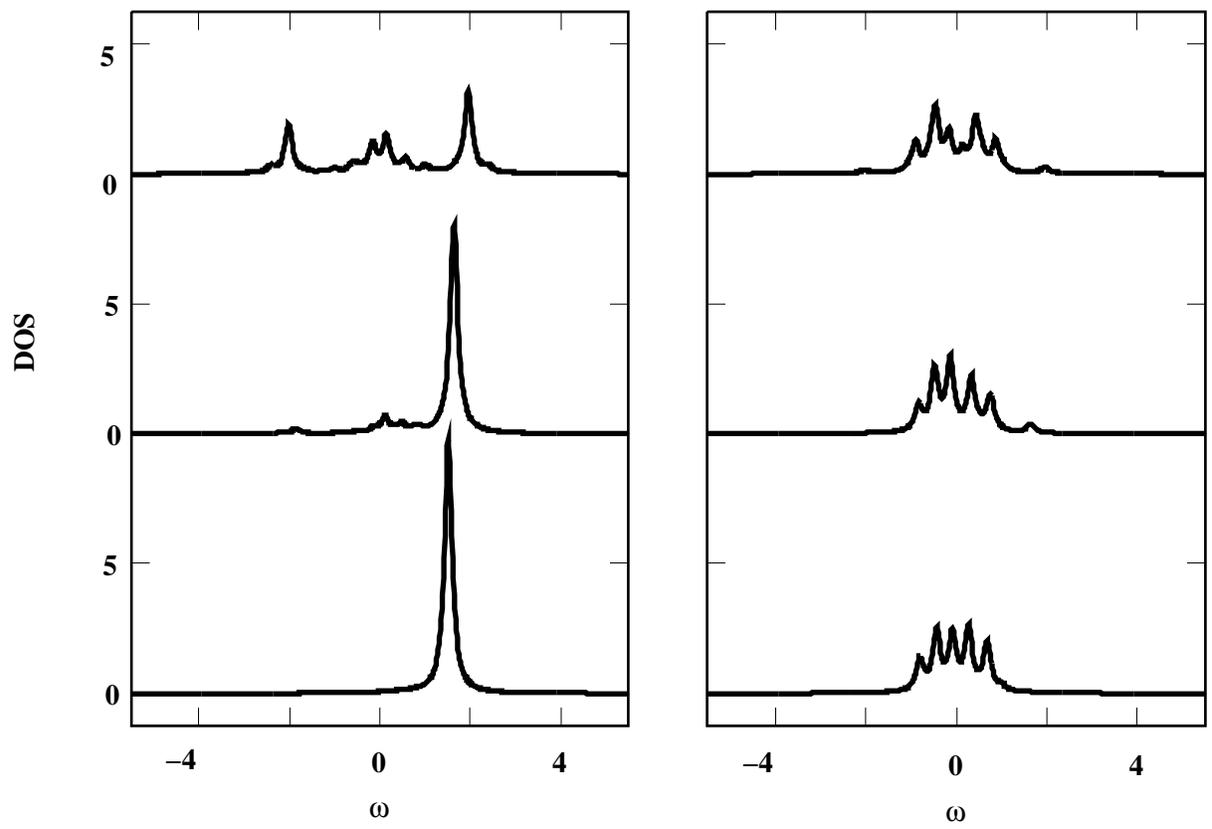
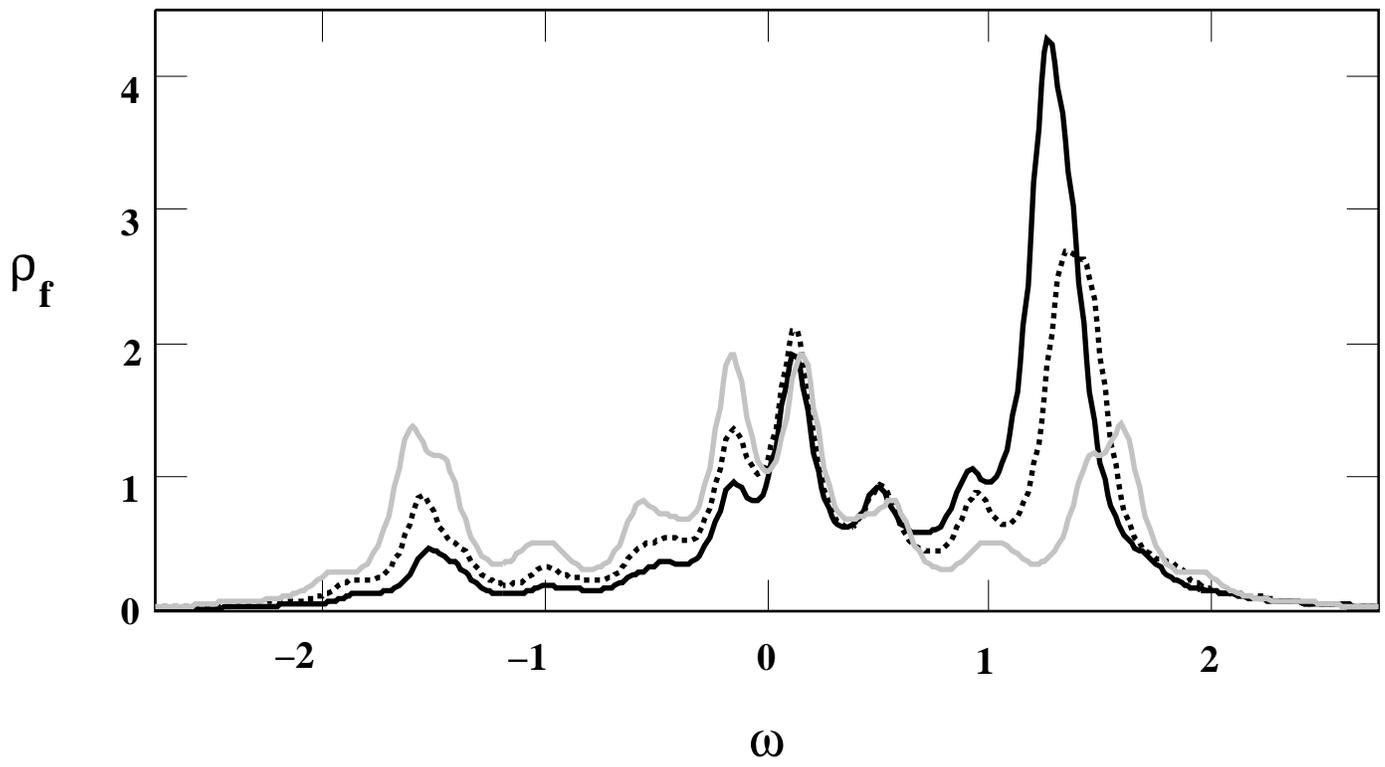
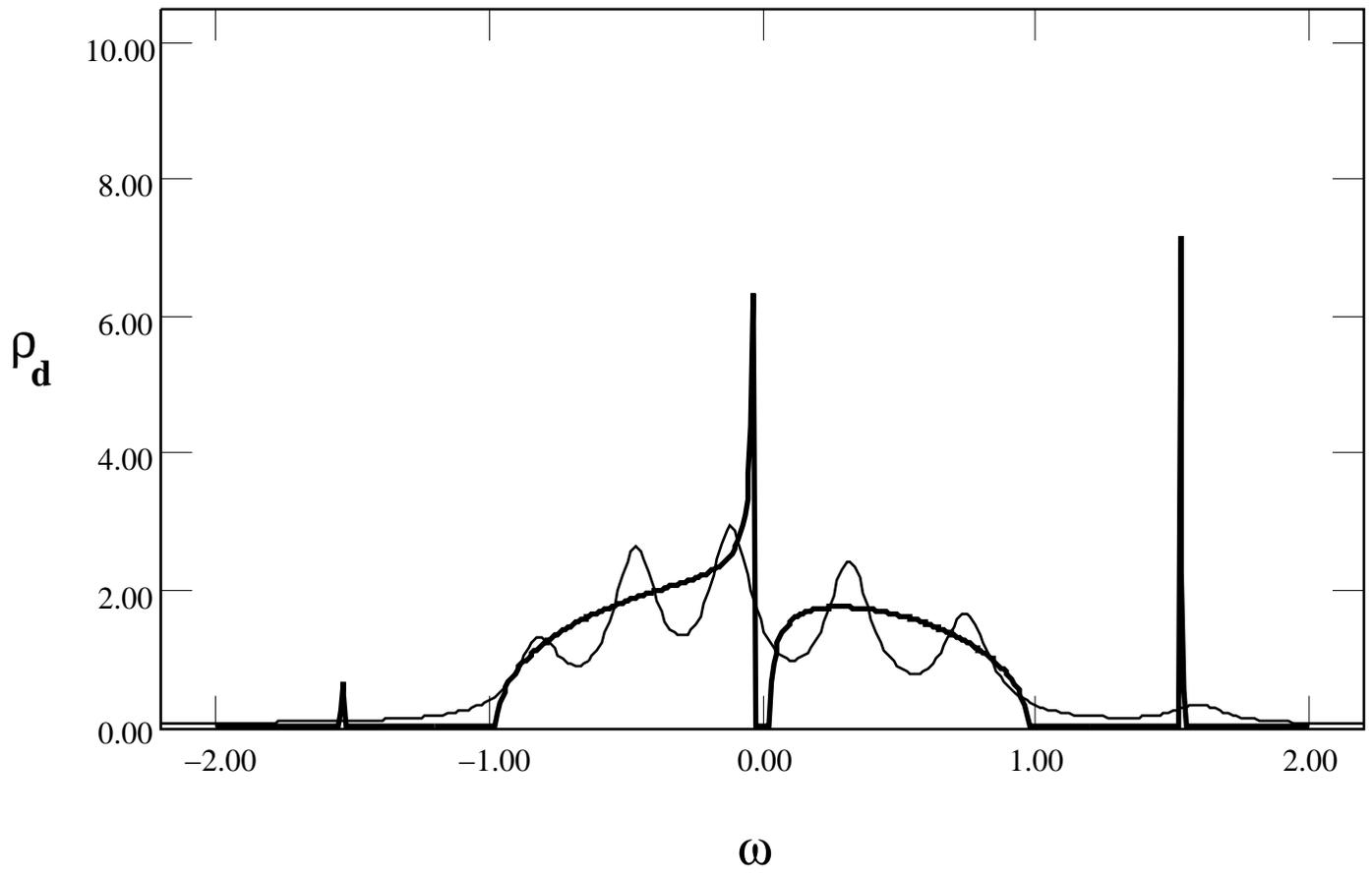
\end{document}